\documentclass{raa} 
\usepackage{graphicx,url,longtable,natbib} %for PS/EPS graphics inclusion, new

\begin{document}

\title{A Proper Motions Study of the Globular Cluster M12 (NGC~6218)}
\volnopage{Vol.0 (200x) No.0, 000--000}
%%preserved for Editor. DOn't remove! \setcounter{page}{1}
%%starting page, preserved for Editor. DOn't remove!
\author{Devesh P. Sariya \inst{1}
\and Ing-Guey Jiang \inst{1}
\and R. K. S. Yadav \inst{2}
 }

\institute{Department of Physics and Institute of Astronomy, 
National Tsing Hua University, 
Hsin-Chu, Taiwan, {\it jiang@phys.nthu.edu.tw};\\
\and Aryabhatta Research Institute of observational sciencES (ARIES), 
Manora Peak Nainital 263 002, India;\\
}

\abstract
{Using astrometric techniques developed by Anderson et al., we determine proper motions (PMs)
in $\sim$14.60$\times$16.53 arcmin$^2$ area of the kinematically ``thick-disk''
globular cluster M12. The cluster's proximity and sparse nature makes it a
suitable target for ground-based telescopes.
Archive images with time gap of $\sim 11.1$ years were observed
with wide-field imager (WFI) mosaic camera mounted on
ESO 2.2 m telescope.
The median value of PM error in both components is $\sim0.7$  mas yr$^{-1}$
for the stars having $V\le20$ mag. PMs are used to determine membership probabilities
and to separate field stars from the cluster sample.
In electronic form, a membership catalog of 3725 stars with precise coordinates,
PMs, $BVRI$ photometry is being provided.
One of the possible applications of the catalog was shown by
gathering the membership information of the variable stars, blue stragglers
and X-ray sources reported earlier in the cluster's region.
}

\authorrunning{Sariya, Jiang \& Yadav } %author_head in even pages
\titlerunning{Proper Motions of M12 stars } % title_head in odd pages
\maketitle

\section{Introduction} 
%\label{sect:intro}
\label{Intro}

Recent studies have brought a new change to the understanding of the
formation and evolution of stars in the globular clusters
showing stars with different chemical composition (e.g. Bedin et al. 2004;
Anderson et al. 2009; Milone et al. 2012; Khamidullina et al. 2014; D'Orazi et al. 2015).
For various reasons, the membership information is useful for evolved stars as well as
fainter stars on the main sequence.
M12 ($\alpha_{2000}$= 16$^{\rm h}$ 47$^{\rm m}$ 14$^{\rm s}$.18,
$\delta_{2000}$= $-01^\circ$ 56$\arcmin$ 54.7$\arcsec$) is a nearby globular cluster
, with the heliocentric distance being 4.8 kpc.
The cluster shows a low-density central region making it an
ideal globular cluster for the ground-based astronomy.
The Galactic location of the cluster places it close to the Galactic disk
$(l, b)$= ($15\fdg72$, $26\fdg31$); and $R_{GC}$= 4.5 kpc.
M12 is a moderately metal-poor (intermediate metallicity, $\rm [Fe/H]=-1.37$) globular cluster.
All the mentioned fundamental parameters of M12 are taken from Harris (1996, updated in 2010).
Giant branch metallicities of M12
show a spread (Zinn \& West 1984; Rutledge et al. 1997).
M12 is a ``second-parameter'' cluster because its horizontal branch (HB)
falls blueward of the RR Lyrae instability strip (Johnson \& Pilachowski 2006).
It has a HB index (Lee et al. 1994) equal to 0.97 (Salaris et al. 2004).
It has been suggested that M12 should belong to the thick disk and not
Galactic halo based on its kinematical properties (Dinescu et al. 1999; Pritzl et al. 2005).

%%%%%%%%%%%%%%%%%%%%%%%%%%%%%%%%%%%%%%%%%%%%%%%%%%%%%%%%%%%%%%%%%%%%%%%%%%%%%%%%%%%%%%%%%%%%%%%%%%%%%%%%%%%%%%%%%%%%%%%%%%%%%%%%%%%%
\begin{figure}
\centering
\includegraphics[width=8.0cm]{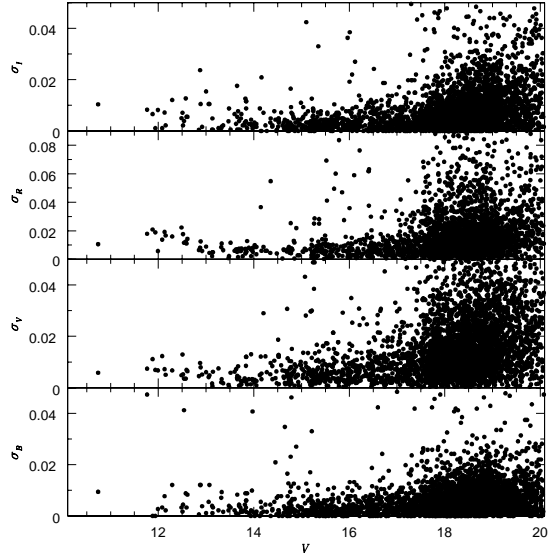}
\caption{Photometric errors in $BVRI$ bands plotted against $V$ magnitude.}
\label{errormag}
\end{figure}
%%%%%%%%%%%%%%%%%%%%%%%%%%%%%%%%%%%%%%%%%%%%%%%%%%%%%%%%%%%%%%%%%%%%%%%%%%%%%%%%%%%%%%%%%%%%%%%%%%%%%%%%%%%%%%%%%%%%%%%%%%%%%%%%%%%%
%%%%%%%%%%%%%%%%%%%%%%%%%%%%%%%%%%%%%%%%%%%%%%%%%%%%%%%%%%%%%%%%%%%%%%%%%%%%%%%%%%%%%%%%%%%%%%%%%%%%%%%%%%%%%%%%%%%%%%%%%%%%%%%%%%%%
\begin{figure}
\centering
\includegraphics[width=8.0cm]{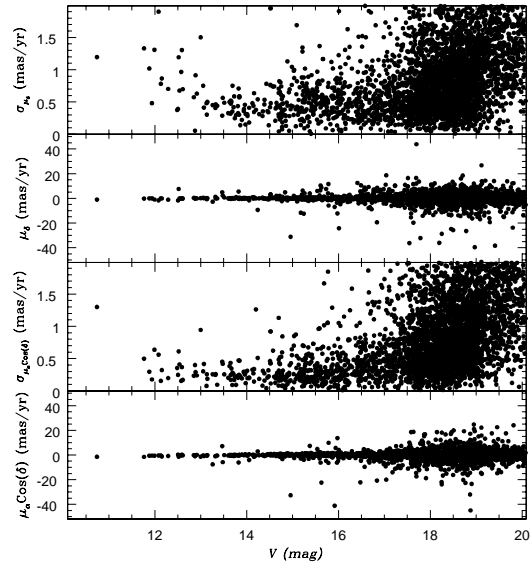}
\caption{Distributions of PMs and their errors as a function of visual magnitude.}
\label{errorpm}
\end{figure}
%%%%%%%%%%%%%%%%%%%%%%%%%%%%%%%%%%%%%%%%%%%%%%%%%%%%%%%%%%%%%%%%%%%%%%%%%%%%%%%%%%%%%%%%%%%%%%%%%%%%%%%%%%%%%%%%%%%%%%%%%%%%%%%%%%%%
Photometry of M12 can be found in various research articles
(e.g. Nassau \& Hynek 1946; Racine 1971; Buonanno et al. 1976;
Mironov et al. 1978; Peikov \& Rusev 1988; Brocato et al. 1996;
Sato et al. 1989; von Braun et al. 2002; Hargis et al. 2004).
von Braun et al. (2002) presented extinction map and reported small differential reddening
across the cluster. The mean value of reddening in the cluster's direction
is $E(B-V)=0.19$ (Harris et al. 1996, 2010 edition).
De Marchi et al. (2006) found that M12 has a very flat mass function and also discussed
the tidal disruption time for this cluster.

Most of the previous studies of variable stars of M12 suggested that M12 is a
variable devoid cluster. Clement et al. (2001) listed only one W Virginis type star
summarizing the variable studies done by that time.
von Braun et al. (2002) found two W Uma type variables.
A recent extensive study by Kaluzny et al. (2015) found 36 variables in M12,
34 of which are new discovery.
A review of M12 variables can be found in Kinman (2016) which also includes
five variables taken from GUVV2 catalog of GALEX far-ultraviolet variables by Wheatley et al. (2008).
Simunovic \& Puzia (2014) present dynamical study of the blue straggler stars in M12
and used radial velocity measurements to separate non-members.
Lu et al. (2009) detected six X-ray sources inside the half-mass radius (2$\farcm$16) of the cluster,
of which two are inside the core radius (0$\farcm$72).
Using {\it Fermi}  large area telescope, Zhang et al. (2016) provided the evidence of
gamma-ray emission from M12.
Pietrukowicz et al. (2008) could not detect any dwarf novae in M12.
M12 is also known to harbor some ``UV bright stars'' (Zinn et al. 1972;
Harris et al. 1983; Geffert et al. 1991).

M12 is particularly appealing for abundance analysis to evaluate
difference between halo and disk cluster systems (Johnson \& Pilachowski 2006).
Klochkova \& Samus (2001) did abundance analysis of three stars
and observed the post AGB star K 413.
Johnson \& Pilachowski (2006) presented chemical abundances and radial velocities
for 21 RGB and asymptotic giant branch stars.
Carretta el al. (2007) determined O and Na abundances as well as Fe abundances for 79
RGB stars from 1 mag below the RGB bump to near the RGB tip.
They also did abundance analysis and found that
Na-O anticorrelation must have been established in the early times of the cluster formation.
D'Orazi et al. (2014) studied Li and Al abundances for a large sample of RGB stars in M12
and found that first-generation and second-generation stars share the same Li content in M12.

Sollima et al. (2016) presented an observational estimate of the fraction and distribution of
dark mass in the innermost region of the cluster.
Lehman \& Scholz (1997) derived the structural parameters of M12 and compared their results with the
values mentioned in literature.
Roederer (2011) found that M12 does not show any evidence of $r$-process dispersion.

Radial velocities of the cluster are studied by
Rastorguev \& Samus (1991); Kiss et al. (2007); Kimmig et al. (2015).
Geffert et al. (1991) provided PMs and membership probabilities of 165 stars, including 13 UV-bright stars
in the cluster region using photographic plates.

Tucholke et al. (1988) presented orbital parameters for M12.
Space motion of the cluster has been studied by various authors
(Brosche et al. 1991; Scholz et al. 1996; Odenkirchen et al. 1997; Dinescu et al. 1999).
The absolute PM of M12 has been listed as
$\mu_{\alpha}cos\delta=1.30\pm0.58$ mas yr$^{-1}$, $\mu_{\delta}=-7.83\pm0.62$ mas yr$^{-1}$
by Dinescu et al. (1999).
A PM study of M12 was conducted by  Zloczewski et al. (2012) (hereafter, Zl12)
which presents relative PMs of stars in the cluster's central region ($\sim$8.82$\times$9.04 arcmin$^2$).
Recently, Narloch et al, (2017) also provided PMs in the region of M12.

It is obvious from the above discussion that M12 serves as an interesting object
for spectroscopy and a membership catalog in the
wider region of the cluster will help in selecting the sample for abundance analysis.
Mosaic CCDs have enabled PM study of star clusters to deeper magnitudes,
in a wider region apart from reducing the required time-gap
(e.g. Anderson et al. 2006; Yadav et al. 2008, 2013; Bellini
et al. 2009; Sariya et al. 2012, 2017a, 2017b; Sariya \& Yadav 2015).
In this study, we conduct a PM study of M12 in a wide region ($\sim$14.60$\times$16.53 arcmin$^2$)
of the cluster's field.

%%%%%%%%%%%%%%%%%%%%%%%%%%%%%%%%%%%%%%%%%%%%%%%%%%%%%%%%%%%%%%%%%%%%%%%%%%%%%%%%%%%%%%%%%%%%%%%%%%%%%%%%%%%%%%%%%%%%%%%%%%%%%%%%%%%%
\begin{table*}
\caption{A brief summary of the archive data observed on 14$^{th}$ May 1999 (first epoch)
and 17$^{th}$ June 2010 (second epoch).}
\centering
\label{log}
\begin{tabular}{ccccc}
\hline
\hline
Epoch & Filters  &  Exposure time & Seeing & Airmass \\
& &(seconds)&(arcsec)& \\
\hline
%\multicolumn{4}{c}{1999 May 14 (First epoch)} \\
First& $B$ &2$\times$240 &0.9 & $\sim$1.1     \\
& $V$ &1$\times$240 &1.0 & $\sim$1.1     \\
& $I$ &2$\times$240 &0.9 & $\sim$1.1     \\
%\multicolumn{4}{c}{ 2010 June 17 (Second Epoch)} \\
Second& $V$ &1$\times$300, 2$\times$250 & 1.5 & $\sim$1.4     \\
& $R$ &3$\times$250               & 1.8 & $\sim$1.5     \\
\hline
\end{tabular}
\end{table*}
%%%%%%%%%%%%%%%%%%%%%%%%%%%%%%%%%%%%%%%%%%%%%%%%%%%%%%%%%%%%%%%%%%%%%%%%%%%%%%%%%%%%%%%%%%%%%%%%%%%%%%%%%%%%%%%%%%%%%%%%%%%%%%%%%%%%
Section~\ref{OBS} discusses the sample of archive data used in this work.
Derivation of the PMs and astrometric and photometric calibration
is performed in Section~\ref{REDUC}.
We discuss the decontamination of field stars in vector point diagrams (VPDs)
and color-magnitude diagrams (CMDs) in Section~\ref{VPD}.
In Section~\ref{MP}, we outline the method to calculate membership probabilities
and compare our results with Zl12.
A context of this work is presented in Section~\ref{app},
where we also briefly summarize the membership catalog.
It is followed by the conclusions in Section~\ref{con}.
%%%%%%%%%%%%%%%%%%%%%%%%%%%%%%%%%%%%%%%%%%%%%%%%%%%%%%%%%%%%%%%%%%%%%%%%%%%%%%%%%%%%%%%%%%%%%%%%%%%%%%%%%%%%%%%%%%%%%%%%%%%%%%%%%%%%
%%%%%%%%%%%%%%%%%%%%%%%%%%%%%%%%%%%%%%%%%%%%%%%%%%%%%%%%%%%%%%%%%%%%%%%%%%%%%%%%%%%%%%%%%%%%%%%%%%%%%%%%%%%%%%%%%%%%%%%%%%%%%%%%%%%%
\begin{figure*}
\centering
\includegraphics[width=\textwidth]{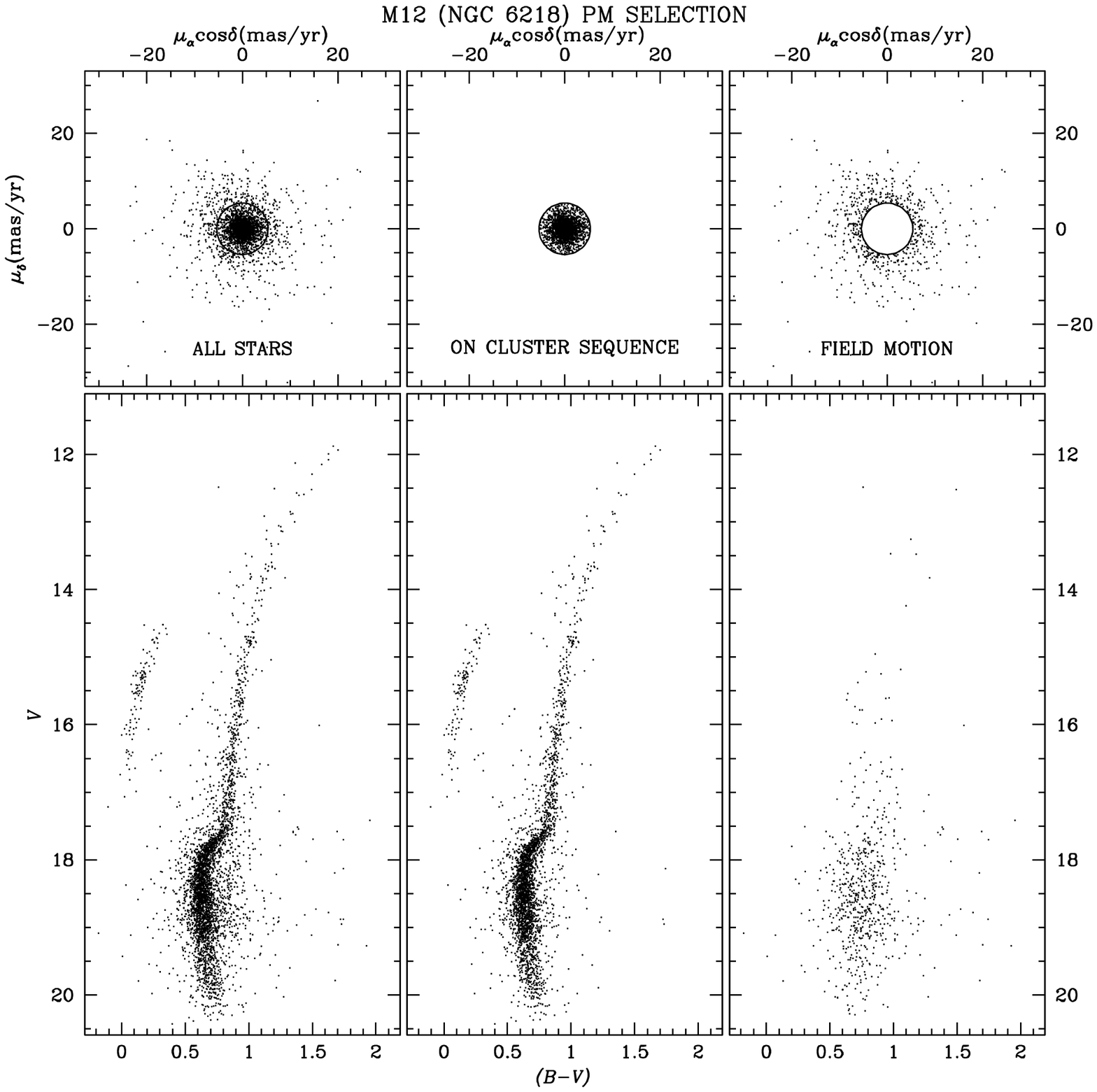}
\caption{
VPDs (top panels) and CMDs (bottom panels) for the studied sample of stars.
The circle in the VPDs shows the provisionally assumed cluster stars.
The left panels show the entire sample of stars.
Using the circle of radius $\sim$5.4 mas~yr$^{-1}$, the central panels
represent the cluster stars and the right panels show the field population.
Only the stars whose PM error does not exceed 2 mas~yr$^{-1}$
in both PM components are considered in making these diagrams.
}
\label{cmd_inst}
\end{figure*}
%%%%%%%%%%%%%%%%%%%%%%%%%%%%%%%%%%%%%%%%%%%%%%%%%%%%%%%%%%%%%%%%%%%%%%%%%%%%%%%%%%%%%%%%%%%%%%%%%%%%%%%%%%%%%%%%%%%%%%%%%%%%%%%%%%%%
%%%%%%%%%%%%%%%%%%%%%%%%%%%%%%%%%%%%%%%%%%%%%%%%%%%%%%%%%%%%%%%%%%%%%%%%%%%%%%%%%%%%%%%%%%%%%%%%%%%%%%%%%%%%%%%%%%%%%%%%%%%%%%%%%%%%
\begin{figure*}
\centering
\includegraphics[width=\textwidth]{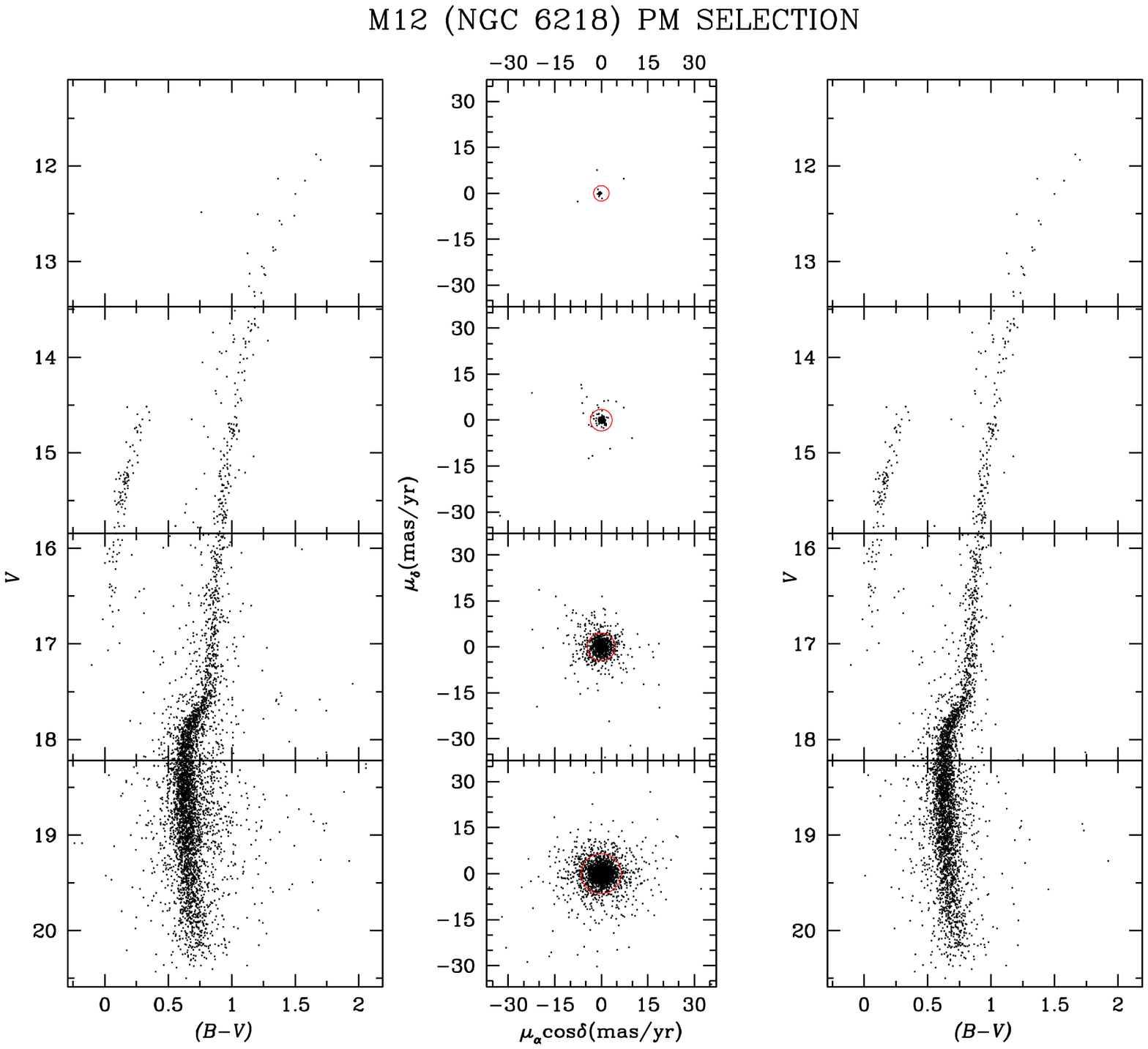}
\caption{Magnitude-binned CMDs of the stars and VPDs according to their
        corresponding magnitude bins.
        The selection criteria are different for different bins.
        PM errors increase from 1.2 mas~yr$^{-1}$ to 2.5 mas~yr$^{-1}$
        from bright to fainter bins.
        The radii of the circles in VPDs increase from 2.5 mas~yr$^{-1}$ from the brightest bin
        to 6.5 mas~yr$^{-1}$ for the faintest bin.
        Left panel CMD shows all the stars plotted in the VPDs.
        While the right panel CMD shows only the stars lying in the circle of corresponding VPDs.
        }
\label{cmd_II}
\end{figure*}
%%%%%%%%%%%%%%%%%%%%%%%%%%%%%%%%%%%%%%%%%%%%%%%%%%%%%%%%%%%%%%%%%%%%%%%%%%%%%%%%%%%%%%%%%%%%%%%%%%%%%%%%%%%%%%%%%%%%%%%%%%%%%%%%%%%%

\section{Sample of Archive Images}
\label{OBS}

The M12 images used in this research are taken from the ESO archive.
The data are observed with 2.2 m ESO/MPI telescope at La Silla, Chile.
This telescope contains a mosaic camera with a combination of 8 EEV CCDs, each having the dimension of 2k$\times$4k pixels.
The images used in this work were obtained between 14$^{th}$ May 1999 and 06$^{th}$ June 2010,
thus providing an epoch gap of $\sim$11.1 years.
The observational log of scientific images with observational area of 34$\times$33 arcmin$^2$ is provided in Table~\ref{log}.
The seeing conditions for the first epoch images are between $\sim$0.9--1.0 arcsec
and airmass values are $\sim$1.1.
The second epoch $V$-band images have seeing values of $\sim1.5$ arcsec and airmass values are $\sim1.4$.
The $R$ band images from the second epoch were not used for the PM work.
They were used only to include $R$-magnitude in our catalog.

\section{Reduction and Calibration of the Data}
\label{REDUC}

The data reduction of the WFI archive images was pursued using the
exclusive astrometric software presented by Anderson et al. (2006, hereafter A06).
Once the images are dealt with initial steps like bias and flat-field corrections,
the Point Spread Function (PSF) is the most crucial factor.
PSF deals with determination of the star positions and fluxes.
Unfortunately, the shape of the PSF changes across the mosaic CCD system, therefore, we can not construct
a single PSF model for the entire CCD systems. A06 designed the program in such a way that the
choice of number of PSFs per chip can
vary from 1 to 15 per CCD chip depending on the richness of the field.
We used an array of 15 PSFs (3$\times$5)
for each CCD chip, thus making 120 total PSFs for the entire mosaic system.
The PSF we use is an $empirical$ PSF represented by a very fine grid.
The fine grid comes from dividing each pixel in 4 equal parts.
Each PSF is extended to a radius of 25 pixels, thus giving 201$\times$201 grid points for every PSF.
The PSF is adjusted so as to be centered on the central gridpoint (101,101).

The second most crucial factor faced by such an analysis is the geometric distortion.
The pixel scale across the wide field-of-view does not remain uniform due to geometric distortion in
WFI@2.2 m. A06 present a 9$\times$17 elements look-up table to account for the geometric distortion.
A06 derived these corrections by observing Galactic bulge in Baade's window with optimally dithered observations.
To achieve the corrections for a given location on the CCD chip, the routine uses a bi-linear interpolation among the four
closest gridpoints. Unfortunately, the distortion correction may not work perfectly owing to the fact that
the distortion correction is less reliable towards the edges of the CCDs as well as the distortion may change with time.

To remove the aforementioned uncertainty in geometric distortion, we use a self-calibration process, called
the local transformation approach. The basic idea of this approach is similar to the classical
``plate-pair method''(e.g., Sanders 1971; Tian et al. 1998). In the local transformation approach,
some local reference stars are used around each target star. Cluster stars are
preferred to be used because they have a lower value of dispersion.
We begin with selecting the stars by making blue and red envelopes to define sequences in the CMD around
the main sequence, sub giant and red giant branch. Further, the stars are selected using PMs and the process is
iterated. Once the locations of the stars are known in all frames, we use a linear transformation
comprising six parameters (three parameters in each of the two coordinates) between frames.
The local transformation  approach can be used for all possible combinations for the first and second epoch images.
Taking average of all possible displacements between different epochs will provide the relative PMs.
Because intra-epoch displacements do not play a role in PMs, they are used to determine PM errors.

\subsection {Astrometric Calibrations}

The astrometric calibration of the pixel positions was carried out using
ESO SKYCAT software.
IRAF\footnote{IRAF is distributed by the
National Optical Astronomical Observatory which is operated by the
Association of Universities for Research in Astronomy, under contact with
the National Science Foundation} tasks $CCMAP$ and $CCTRAN$ were used for this
aim and the rms of the transformations was found to be $\sim40$ mas.
A single plate model consisting of linear and quadratic terms
and a small but significant cubic term was used.
The precision in defining distortion correction and
the stability of intra-chip positions made it possible to use this single plate model.
This model also gets rid of the impressions generated by the differential refraction.

\subsection{Photometric Calibrations}

To convert the instrumental to standard magnitudes we used the secondary standard
stars provided by
P. Stetson\footnote{http://www3.cadc-ccda.hia-iha.nrc-cnrc.gc.ca/community/STETSON/standards/}.
The number of secondary standard stars used for transformations are 67, 66, 44 and 62 in $B$, $V$, $R$ and $I$
filters. The magnitudes and colors range of standard stars are
15$\le V \le$19.5, 0.1$\le (B-V)\le$1.3, 0.3$\le (V-R)\le$0.6 and 0.2$\le (V-I)\le$1.5.

The transformation equations used to calibrate the magnitudes are

\begin{center}
$ B_{\rm std} = B_{\rm ins} + C_b\times(B_{\rm ins} - V_{\rm ins}) + Z_b $

$ V_{\rm std} = V_{\rm ins} + C_v\times(B_{\rm ins} - V_{\rm ins}) + Z_v $

$ R_{\rm std} = R_{\rm ins} + C_r\times(V_{\rm ins} - R_{\rm ins}) + Z_r $

$ I_{\rm std} = I_{\rm ins} + C_i\times(V_{\rm ins} - I_{\rm ins}) + Z_i $,\\
\end{center}

where, subscripts ``ins'' and ``std'' represent the instrumental and standard magnitudes. $Z_b, Z_v, Z_r$ and $Z_i$ represent the zeropoints in $B$, $V$, $R$ and $I$ filters respectively.
The color term values are 0.39, $-$0.11, 0.03 and 0.10 while zeropoints are 24.77, 23.83, 24.10
and 23.27 for  $B$, $V$ $R$ and $I$ filters respectively.
The present values are very close to the values mentioned on the WFI@2.2 m
webpage\footnote{www.eso.org/sci/facilities/lasilla/instruments/wfi/inst/zeropoints.html}.

The photometric rms with calibrated $V$ magnitudes is shown in Fig.~\ref{errormag}.
A few brighter stars have more errors because of saturation and crowding. 
The bright star at $V\sim10.7$ mag is located towards the CCD edge in our catalogue
where crowding is not a problem. So, it has relatively less photometric errors. 
Photometric errors for saturated stars located in the central region of the cluster are higher. 
However, since the distortion solution is poor towards the outer region for this CCD imager,
this bright star  ($V\sim10.7$) has larger PM errors as shown in Fig.~\ref{errorpm}. 
The photometric standard deviations for individual photometric bands were determined using multiple observations,
all reduced to a common reference frame.
For $V<18$ mag, the mean values of errors are $\sim$0.005 mag for
$B$ and $I$ filters, $\sim$0.01 mag for $V$ band and $\sim$0.07 mag for the $R$ band magnitudes.

\section {M12 PMs, VPD and CMD}
\label{VPD}

To derive the PMs of the stars in the cluster region, we need at least two epochs data. For this
cluster M12, we have CCD data available in 1999 and 2010 epoch. The time baseline of $\sim$ 11.1 years
for this cluster is
suitable for the separation of cluster stars from field stars. 
Five images in the first epoch and
three images in the second epoch in $BVI$ filters are used to determine PMs. 
Due to poor image quality, R filter images from the
2010 epoch were not used during the PM calculation process.
To transform the coordinates from one epoch to second epoch, cluster sequence stars have been
used as a first guess. These selected stars have $V$ magnitudes range between 13 and 19 mag
with PM errors  $<$2 mas~yr$^{-1}$. To minimize the effect of uncorrected distortion
residuals, a local transformation approach based on the closest 25 stars on the same CCD was
applied. No systematics larger than random errors near the
corners or edges of the CCD chips were found.
The PM calculating routine was iterated subsequently and
it helped in removing some stars from the initial photometric members list. Even if the colors of some stars
were such as to place them on the fiducial cluster sequences in the CMD,
they were removed from the preliminary members list
if their PMs were inconsistent to make them fit for cluster membership.

Figure~\ref{errorpm} represents the distribution of PMs in both $X$ and $Y$directions and their standard errors
as a function of $V$ magnitude. The median values of PM error is $\sim$0.7 mas~yr$^{-1}$ for stars
brighter than $V\sim$20 mag.

The underlying beauty of PMs in a globular cluster is to distinguish the cluster sample from the field stars,
thereby, producing a CMD with most likely cluster members. This is evident in Fig.~\ref{cmd_inst}.
PMs in both directions are plotted as the vector point diagrams (VPDs) in the top panels of the figure.
The respective $(B-V), V$ CMDs are shown in the bottom panels of Fig.~\ref{cmd_inst}.
In the left panels of this figure, the entire sample of stars studied here is plotted.
In the middle panels, the preliminarily assumed cluster
members are shown, while the right panels show the assumed field stars.
In the top panels of the VPD, a circle of radius $\sim$5.4 mas~yr$^{-1}$ is
shown around the centroid of PMs of stars.
This serves as a provisional criterion to decide on the cluster member stars.
This radius is chosen in such a way to include maximum possible cluster members while
noticing its effect on the CMD in decontaminating the field stars.
However, some cluster members having poorly measured PMs can be lost and some field stars
sharing their motion with M12 may get included as preliminary members.
The fact that the distribution of cluster members' PM is round, gives the implication that
the present PM analysis is not influenced by any sort of systematics.

Magnitude-binned $V$ versus $(B-V)$ CMDs and corresponding VPDs are plotted in Figure~\ref{cmd_II}.
The point to note here is that we adopted different selection criteria for different magnitude bins
such as the criterion becomes less restrictive towards the fainter magnitude bins.
The middle panels show the PM VPDs while the binned-CMD of likely cluster members is shown in the right panels.
The value of PM errors for the bins increase from $\sim$1.2 mas~yr$^{-1}$ to $\sim$2.5 mas~yr$^{-1}$ from brightest down to faintest bins.
Likewise, the radii of the circle in the VPDs vary from $\sim$2.5 mas~yr$^{-1}$ to $\sim$6.5 mas~yr$^{-1}$ in the same order.
It is evident that we obtained a good separation of field stars up to $\sim$ 18 mag and the measurement
errors take over after that.

\section{The Cluster Membership}
\label{MP}

Proper motions are frequently used to determine kinematical membership probabilities.
In our study of M12, we used the membership calculation method presented by Balaguer-N\'{u}\~{n}ez et al. (1998).
This method has been previously used for WFI data of star clusters
(Bellini et al. 2009, Sariya et al. 2012, 2017a; Yadav et al. 2013; Sariya \& Yadav 2015).
The membership in this method is determined by the superposition of two different frequency distribution functions.
For the cluster and field star distributions, two different distribution functions ($\phi_c^{\nu}$) and ($\phi_f^{\nu}$)
are constructed for a particular i$^{th}$ star.
%considering it a cluster star and field star respectively.
Neglecting the spatial distribution of the stars,
the values of frequency distribution functions are given as follows:

\begin{center}
   $\phi_c^{\nu} =\frac{1}{2\pi\sqrt{{(\sigma_c^2 + \epsilon_{xi}^2 )} {(\sigma_c^2 + \epsilon_{yi}^2 )}}}$

$\times$ exp$\{{-\frac{1}{2}[\frac{(\mu_{xi} - \mu_{xc})^2}{\sigma_c^2 + \epsilon_{xi}^2 } + \frac{(\mu_{yi} - \mu_{yc})^2}{\sigma_c^2 + \epsilon_{yi}^2}] }\}$ \\
\end{center}
%\end{equation}
\begin{center}
and\\
\end{center}
%\begin{equation}
\begin{center}
$\phi_f^{\nu} =\frac{1}{2\pi\sqrt{(1-\gamma^2)}\sqrt{{(\sigma_{xf}^2 + \epsilon_{xi}^2 )} {(\sigma_{yf}^2 + \epsilon_{yi}^2 )}}}$

$\times$ exp$\{{-\frac{1}{2(1-\gamma^2)}[\frac{(\mu_{xi} - \mu_{xf})^2}{\sigma_{xf}^2 + \epsilon_{xi}^2}}
-\frac{2\gamma(\mu_{xi} - \mu_{xf})(\mu_{yi} - \mu_{yf})} {\sqrt{(\sigma_{xf}^2 + \epsilon_{xi}^2 ) (\sigma_{yf}^2 + \epsilon_{yi}^2 )}} + \frac{(\mu_{yi} - \mu_{yf})^2}{\sigma_{yf}^2 + \epsilon_{yi}^2}]\}$\\
\end{center}

where ($\mu_{xi}$, $\mu_{yi}$) are the PMs of $i^{th}$ star.
The PM errors are represented by ($\epsilon_{xi}$, $\epsilon_{yi}$).
The cluster's PM center is given by
($\mu_{xc}$, $\mu_{yc}$)
and ($\mu_{xf}$, $\mu_{yf}$) represent the center of field PM values.
The intrinsic PM dispersion for the cluster stars is denoted by
$\sigma_c$, whereas $\sigma_{xf}$ and $\sigma_{yf}$ provide the
intrinsic PM dispersions for the field populations. The correlation coefficient $\gamma$ is calculated as:\\

%\begin{equation}
\begin{center}
$\gamma = \frac{(\mu_{xi} - \mu_{xf})(\mu_{yi} - \mu_{yf})}{\sigma_{xf}\sigma_{yf}}$.
\end{center}

Only the stars with PM errors $\le$2 mas~yr$^{-1}$ are used in calculating
$\phi_c^{\nu}$ and $\phi_f^{\nu}$.
As expected from the VPD, the center of the cluster stars is found to be
($\mu_{xc}$, $\mu_{yc}$)=(0, 0) mas~yr$^{-1}$.
The  intrinsic PM dispersion for the cluster stars ($\sigma_c$)
could not be determined reliably using our PM data.
Kimmig et al. (2015) compared their results with the literature values for radial velocity data and
listed the best fit value of radial velocity dispersion for M12
as 4.3 km~sec$^{-1}$. Considering the value of distance of M12 as
4.8 kpc (Harris 1996, 2010 edition),
the internal PM dispersion  ($\sigma_c$) becomes
$\sim$0.19 mas~yr$^{-1}$.
For the field population, we have
($\mu_{xf}$, $\mu_{yf}$) = ($-$5.9, 5.8) mas yr$^{-1}$ and
($\sigma_{xf}$, $\sigma_{yf}$) = (3.6, 3.9) mas yr$^{-1}$. 
The mentioned field parameters are determined considering the
field stars lying on the top left corner of the VPDs (Fig.~\ref{cmd_inst}) as the field stars
because the field motion is more significant towards that region.  
This selection is further justified by looking at the magnitude-binned panels
in Fig.~\ref{cmd_II} where field stars upto $V\sim18.2$ magnitude show their 
clear distribution in the chosen direction. The stars fainter than
 $V\sim18.2$ do not have much clear distribution in a particular direction 
due to increasing PM errors.

Considering the normalized numbers of cluster stars and field stars as
$n_{c}$ and $n_{f}$ respectively
(i.e., $n_c + n_f = 1$), the total distribution function can be calculated as:\\

\begin{center}
$\phi = (n_{c}~\times~\phi_c^{\nu}) + (n_f~\times~\phi_f^{\nu})$,  \\
\end{center}

As a result, the membership probability for the $i^{th}$ star is given by:\\
\begin{center}
$P_{\mu}(i) = \frac{\phi_{c}(i)}{\phi(i)}$. \\
\end{center}

%%%%%%%%%%%%%%%%%%%%%%%%%%%%%%%%%%%%%%%%%%%%%%%%%%%%%%%%%%%%%%%%%%%%%%%%%%%%%%%%%%%%%%%%%%%%%%%%%%%%%%%%%%%%%%%%%%%%%%%%%%%%%%%%%%%
\begin{figure}
\vspace{-3.0cm}
\centering
\includegraphics[width=8.0cm]{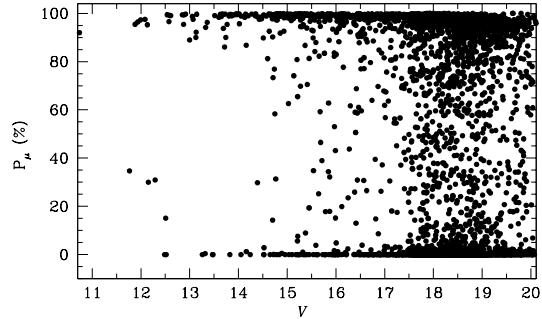}
\caption{Distribution of the cluster membership probabilities with $V$ magnitude.
}
\label{VvsMP}
\end{figure}
%%%%%%%%%%%%%%%%%%%%%%%%%%%%%%%%%%%%%%%%%%%%%%%%%%%%%%%%%%%%%%%%%%%%%%%%%%%%%%%%%%%%%%%%%%%%%%%%%%%%%%%%%%%%%%%%%%%%%%%%%%%%%%%%%%%
%%%%%%%%%%%%%%%%%%%%%%%%%%%%%%%%%%%%%%%%%%%%%%%%%%%%%%%%%%%%%%%%%%%%%%%%%%%%%%%%%%%%%%%%%%%%%%%%%%%%%%%%%%%%%%%%%%%%%%%%%%%%%%%%%%%
\begin{figure}
\centering
\includegraphics[width=8.0cm]{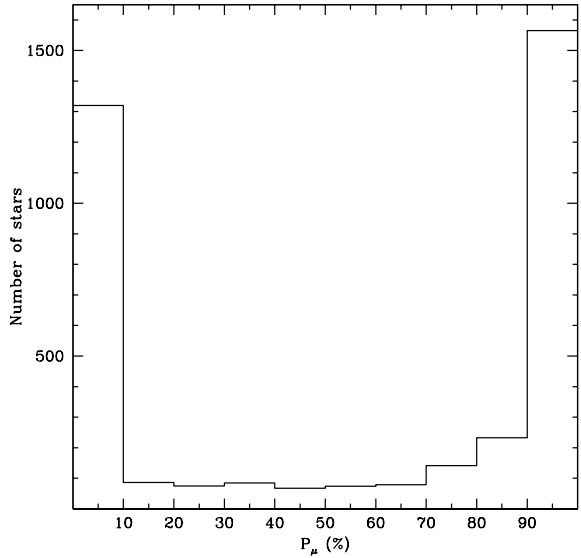}
\caption{Histogram of the membership probabilities.}
\label{histmp}
\end{figure}
%%%%%%%%%%%%%%%%%%%%%%%%%%%%%%%%%%%%%%%%%%%%%%%%%%%%%%%%%%%%%%%%%%%%%%%%%%%%%%%%%%%%%%%%%%%%%%%%%%%%%%%%%%%%%%%%%%%%%%%%%%%%%%%%%%%
%%%%%%%%%%%%%%%%%%%%%%%%%%%%%%%%%%%%%%%%%%%%%%%%%%%%%%%%%%%%%%%%%%%%%%%%%%%%%%%%%%%%%%%%%%%%%%%%%%%%%%%%%%%%%%%%%%%%%%%%%%%%%%%%%%%
\begin{figure}
\centering
\includegraphics[width=8.0cm]{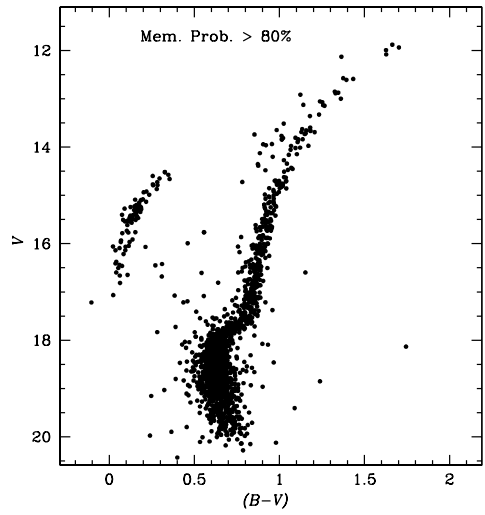}
\caption{CMD showing only those stars for which membership probability value is greater than 80\%.}
\label{mp_cmd}
\end{figure}
%%%%%%%%%%%%%%%%%%%%%%%%%%%%%%%%%%%%%%%%%%%%%%%%%%%%%%%%%%%%%%%%%%%%%%%%%%%%%%%%%%%%%%%%%%%%%%%%%%%%%%%%%%%%%%%%%%%%%%%%%%%%%%%%%%%%
%%%%%%%%%%%%%%%%%%%%%%%%%%%%%%%%%%%%%%%%%%%%%%%%%%%%%%%%%%%%%%%%%%%%%%%%%%%%%%%%%%%%%%%%%%%%%%%%%%%%%%%%%%%%%%%%%%%%%%%%%%%%%%%%%%%
\begin{figure}
\centering
\includegraphics[width=8.0cm]{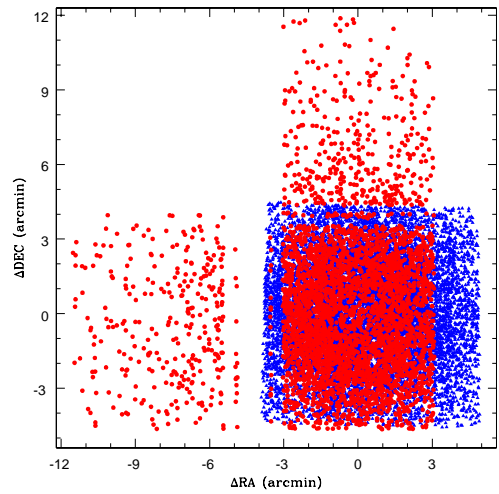}
\caption{Comparison of the spatial distributions of stars in
our study (red filled circles) and Zl12 (blue triangles). The differences in R.A. and dec. are plotted here
between individual stars' coordinates and that of the center of M12 in
arcmin units.
}
\label{spatialboth}
\end{figure}
%%%%%%%%%%%%%%%%%%%%%%%%%%%%%%%%%%%%%%%%%%%%%%%%%%%%%%%%%%%%%%%%%%%%%%%%%%%%%%%%%%%%%%%%%%%%%%%%%%%%%%%%%%%%%%%%%%%%%%%%%%%%%%%%%%%%
%%%%%%%%%%%%%%%%%%%%%%%%%%%%%%%%%%%%%%%%%%%%%%%%%%%%%%%%%%%%%%%%%%%%%%%%%%%%%%%%%%%%%%%%%%%%%%%%%%%%%%%%%%%%%%%%%%%%%%%%%%%%%%%%%%%
\begin{figure}
\centering
\includegraphics[width=8.0cm]{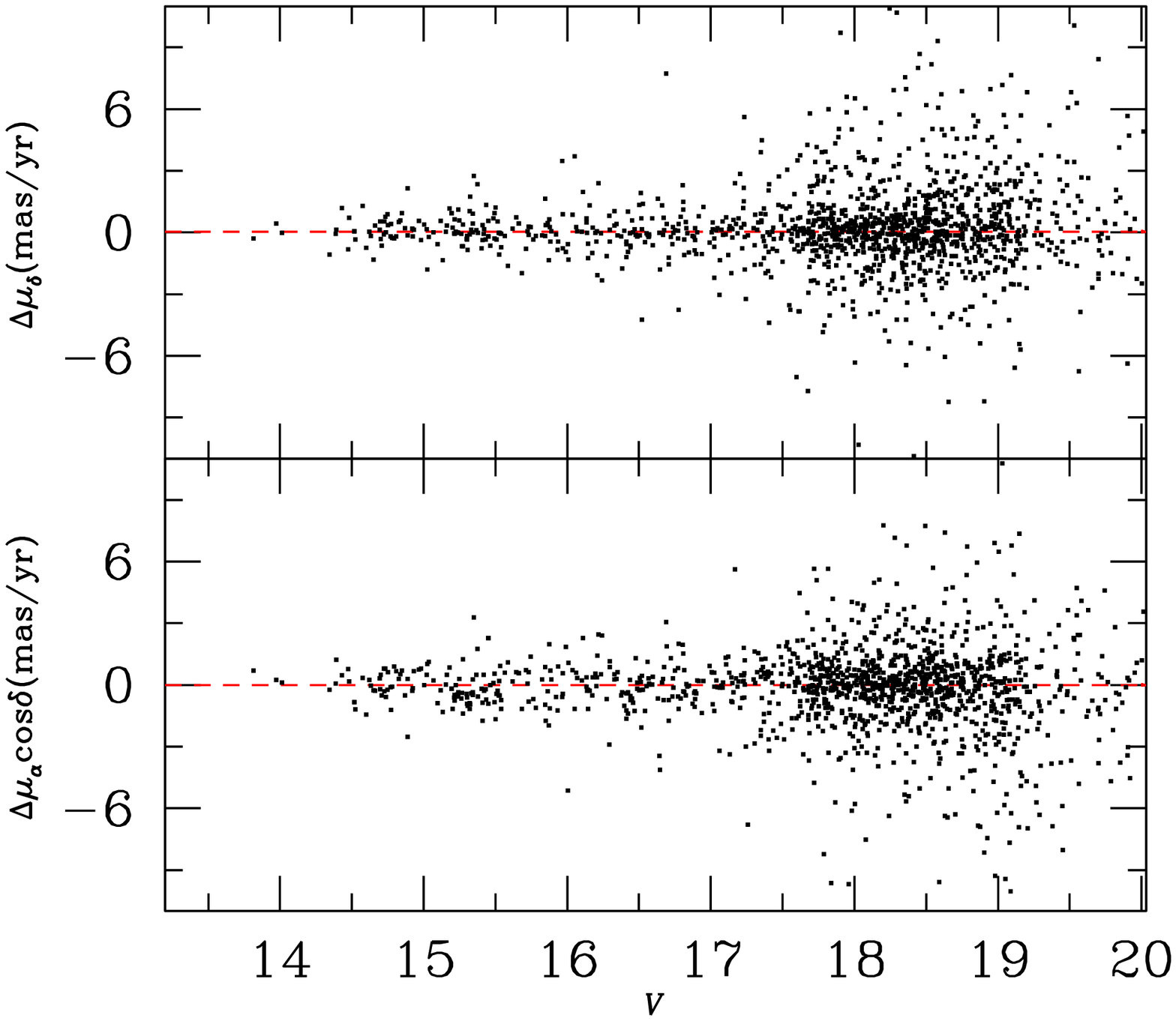}
\caption{Plot of the PM differences between Z12 catalog and the present study vs $V$ magnitude.
3$\sigma$-clipped median PM differences in both PM components are shown by two horizontal dashed lines.}
\label{kamil}
\end{figure}
%%%%%%%%%%%%%%%%%%%%%%%%%%%%%%%%%%%%%%%%%%%%%%%%%%%%%%%%%%%%%%%%%%%%%%%%%%%%%%%%%%%%%%%%%%%%%%%%%%%%%%%%%%%%%%%%%%%%%%%%%%%%%%%%%%%
%%%%%%%%%%%%%%%%%%%%%%%%%%%%%%%%%%%%%%%%%%%%%%%%%%%%%%%%%%%%%%%%%%%%%%%%%%%%%%%%%%%%%%%%%%%%%%%%%%%%%%%%%%%%%%%%%%%%%%%%%%%%%%%%%%%%

A plot of cluster membership probabilities with $V$ magnitude is shown in Figure~\ref{VvsMP}.
It is evident from the figure that the cluster stars and field stars show
separation as two separate distributions of stars near the membership values
$P_{\mu}\sim$100$\%$ and  $P_{\mu}\sim$0$\%$.
For the stars fainter than $V\sim$18 mag, the increasing measurement errors make the separation ambiguous
with more stars falling at intermediate values of membership probabilities.
For these stars, the mean values of P$_{\mu}$ for cluster members is falling,
while it is rising for the field stars.
A histogram of the membership probabilities is shown in Fig.~\ref{histmp}.
Total number of stars shown in the histogram are 3725. We have
1320 stars with membership probability value less than 10\%.
The most probable cluster members, stars with $P_{\mu}\ge$90$\%$ are 1565 in number.

CMD of 1798 stars having membership probabilities $>80\%$ is shown in Fig.~\ref{mp_cmd}.
This CMD exhibits clear cluster sequences for stars brighter than $V \sim20$ mag.
Also, this CMD shows stars of various evolutionary stages like sub-giants, red giants, horizontal branch
stars and blue stragglers.
All the cluster sequences in this CMD look cleaner with minimal field star contamination.
M12 is known to have a CMD very similar to
another globular cluster M10 (von Braun et al. 2002). Also, the CMD exhibits an extremely blue horizontal branch.
Jasniewicz et al. (2004) list the HB morphology parameter (Buonanno et al. 1997) of this cluster as 0.80.

\subsection{Comparison with Zl12}

We compared the results of the present work with the catalog provided by Zl12.
The comparison was done in the spatial distribution of the stars as well as between the PMs of both catalogs.
Figure~\ref{spatialboth} shows the spatial distributions of both catalogs with different colors.
It is evident that our catalog (red filled circles) covers a wider region of the cluster than Zl12 (blue triangles).
We have PM information for about 662 stars in the additional area we studied than the field covered by Zl12.
The differences in PMs in both directions for the common stars between the two catalogs is
presented in Fig.~\ref{kamil}. The red dotted lines in the figure show the
3$\sigma$-clipped median of the PM differences as
$0.003 (\sigma = 0.683)$ mas yr$^{-1}$ and $0.039 (\sigma = 0.587)$ mas yr$^{-1}$.
Our PMs are clearly in consistence with those presented by Zl12 up to 20 $V$ mag.

%%%%%%%%%%%%%%%%%%%%%%%%%%%%%%%%%%%%%%%%%%%%%%%%%%%%%%%%%%%%%%%%%%%%%%%%%%%%%%%%%%%%%%%%%%%%%%%%%%%%%%%%%%%%%%%%%%%%%%%%
\begin{figure}
\centering
\includegraphics[width=8.0cm]{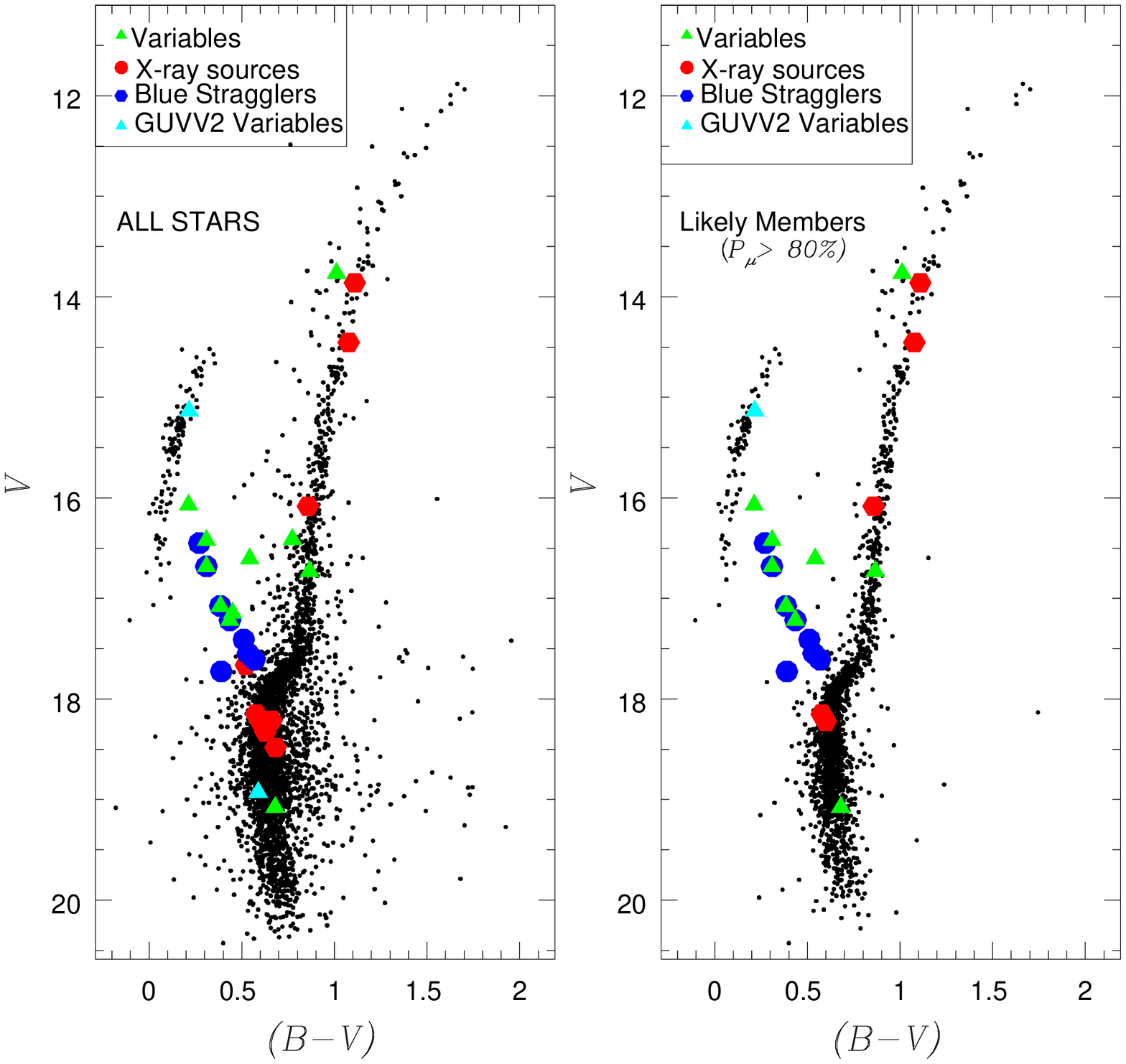}
\caption{
{\em (Left:)} The CMD of all stars in our membership catalog showing the position of
variable stars, blue stragglers and X-ray sources. These peculiar sources are shown with
different symbols as indicated in the inset.
{\em (Right:)}  The same CMD showing only those stars which have a membership probability value higher than 80\%.
}
\label{cmd_vari}
\end{figure}
%%%%%%%%%%%%%%%%%%%%%%%%%%%%%%%%%%%%%%%%%%%%%%%%%%%%%%%%%%%%%%%%%%%%%%%%%%%%%%%%%%%%%%%%%%%%%%%%%%%%%%%%%%%%%%%%%%%%%%%%%%%%%%%%%%%
%%%%%%%%%%%%%%%%%%%%%%%%%%%%%%%%%%%%%%%%%%%%%%%%%%%%%%%%%%%%%%%%%%%%%%%%%%%%%%%%%%%%%%%%%%%%%%%%%%%%%%%%%%%%%%%%%%%%%%%%%%%%%%%%%%%
\begin{table*}
\caption{Cluster membership status  of the variables (ID$_V$) listed in Kaluzny et al. (2015)
and the UV variables found in GUVV2 catalog (ID$_{V(UV)}$. ID represents the corresponding
star number in our catalog.}
\centering
%\tiny
\begin{tabular}{ccc|ccc}
\hline
& Variables  &     &   &  GUVV2 Variables   &         \\
\hline
ID$_V$  &  P$_{\mu}$   & ID        &  ID$_{V(UV)}$  &  P$_{\mu}$    & ID \\
&(\%) & & &(\%) &\\
\hline
V6   &  97.03  & 3561  & 334  &  99.11  &  695  \\
V8   &  98.24  &  460  & 336  &  58.15  & 3022  \\
V12  &  96.97  &  760  &      &      &       \\
V18  &  00.00  &  428  &      &      &       \\
V22  &  99.04  &  769  &      &      &       \\
V23  &  99.22  &  410  &      &      &       \\
V24  &  00.54  &  985  &      &      &       \\
V25  &  99.04  &  417  &      &      &       \\
V28  &  95.40  &  642  &      &      &       \\
V29  &  99.39  &  133  &      &      &       \\
V36  &  98.70  &  426  &      &      &       \\
\hline
\label{var01}
\end{tabular}
\end{table*}
%%%%%%%%%%%%%%%%%%%%%%%%%%%%%%%%%%%%%%%%%%%%%%%%%%%%%%%%%%%%%%%%%%%%%%%%%%%%%%%%%%%%%%%%%%%%%%%%%%%%%%%%%%%%%%%%%%%%%%%%%%%%%%%%%%%
\begin{table*}
\caption{Membership probability for
X-ray sources (ID$_X)$ given by Lu et al. (2009)
and blue stragglers (ID$_B$) provided by Simunovic \& Puzia (2014). The corresponding
star number in our catalog is shown under the column of ID.}
\centering
%\tiny
\begin{tabular}{ccc|ccc}
\hline
& X-ray sources  &     &   &   Blue Stragglers   &         \\
\hline
ID$_X$  &  P$_{\mu}$   & ID        &  ID$_B$  &  P$_{\mu}$    & ID \\
&(\%) & & &(\%) &\\
\hline
 CX1  &  98.97  &  570  & BSS1   &  95.96  &  381   \\
 CX2  &  99.27  &  548  & BSS3   &  99.22  &  410   \\
 CX3  &  00.00  & 1663  & BSS9   &  90.88  &  565   \\
 CX4  &  98.48  &  478  & BSS19  &  92.28  &  238   \\
 CX8  &  99.44  & 1676  & BSS20  &  98.38  &  299   \\
 CX9  &  41.14  & 1957  & BSS22  &  99.04  &  417   \\
 CX10 &  00.09  & 2186  & BSS27  &  81.89  &  551   \\
 CX12 &  00.09  & 1477  & BSS30  &  99.04  &  769   \\
 CX18 &  98.21  & 1922  &        &         &        \\
\hline
\label{var02}
\end{tabular}
\end{table*}
%%%%%%%%%%%%%%%%%%%%%%%%%%%%%%%%%%%%%%%%%%%%%%%%%%%%%%%%%%%%%%%%%%%%%%%%%%%%%%%%%%%%%%%%%%%%%%%%%%%%%%%%%%%%%%%%%%%%%%%%%%%%%%%%%%%

\section{Discussion}
\label{app}
\subsection{Membership of Peculiar Stars}

The credit to search most of the variables of M12 goes to the extensive study by
Kaluzny et al. (2015). Including two already discovered variables, they reported 36 variables in the region of M12.
We found 11 variables out of these and list their membership status in Table~\ref{var01}.
It can be inferred from the table that except V18 and V24 for which $P_{\mu}\sim$0$\%$, all the other discovered variables
(V6, V8, V12, V22, V23, V25, V28, V29 and V36) are most likely cluster members.
Kaluzny et al. (2015) also mentioned the membership status of these variables based on the Zl12 catalog.
The membership status determined by us agrees with Kaluzny et al. (2015) for
all the variables except V24.
Wheatley et al. (2008) provided a catalog (GUVV2) of far-ultraviolet variables using $GALEX$.
Kinman (2016) presented a list of 4 far-UV variables in the cluster region using GUVV2 catalog.
Their membership status is given in Table~\ref{var01} where it can be seen that one of the far-UV variables
( ID$_{V(UV)}$=334) belongs to the cluster.
Lu et al. (2009) detected 20 X-ray sources using {\it Chandra X-ray observatory} data in the direction of M12.
Their membership status is shown in Table~\ref{var02}. Among the X-ray counterparts we found in our catalog,
5 sources (CX1, CX2, CX4, CX8, CX18) have their membership probability  $P_{\mu}>$98$\%$.
They are most likely cluster members. Three X-ray sources (CX3, CX10 and CX12) have zero membership probability.
So, they should not be the cluster members.
Simunovic \& Puzia (2014) carried out medium-resolution spectroscopy
of the cluster and reported some blue stragglers in the cluster. As can be seen in Table~\ref{var02},
all of the 8 blue stragglers we found are having $P_{\mu}>$81$\%$. They all should be belonging to M12 according to our catalog.
Figure~\ref{cmd_vari} shows all the above mentioned sources in the CMD of M12.
CMD on the left shows all the objects listed in Tables~\ref{var01} and~\ref{var02}, whereas
the CMD in the right panel is plotted with only the stars with membership probabilities $>$80\%.
Variables, blue stragglers and X-ray sources shown in the right panel are most likely cluster members.

\subsection{The Electronic Catalog}
\label{catl}

The electronic catalog provides photometric and kinematical data for 3725 stars in the region of globular cluster M12.
The catalog includes relative PMs, PM errors, membership probabilities,
$B, V, R, I$ photometry with rms errors. For a few stars, $R$ band photometry is missing.
For those stars, a flag equal to 99.9999 for the magnitude and 0.9999 for the error is set.
A small sample from the electronic catalog along with the header information is shown in Table \ref{cata}.

%%%%%%%%%%%%%%%%%%%%%%%%%%%%%%%%%%%%%%%%%%%%%%%%%%%%%%%%%%%%%%%%%%%%%%%%%%%%%%%%%%%%%%%%%%%%%%%%%%%%%%%%%%%%%%%%%%%%%%%%%%%%%%%%%%%
%\begin{sidewaystable*}
\begin{table*}
\caption{Some initial lines from the electronic membership catalog for M12.}
\centering
\vspace{0.4cm}
%\tiny
\resizebox{\textwidth}{!}{
\begin{tabular}{cccccccccccccccccc}
\hline\hline
ID & $\alpha_{2000}$ & $\delta_{2000}$ &X&Y & $\mu_{\alpha}cos(\delta)$ &$ \sigma_{\mu_{\alpha}cos(\delta)}$ &$\mu_{\delta}$ & $\sigma_{\mu_{\delta}}$&B&$\sigma_B$&V&$\sigma_V$&$R$&$\sigma_R$&I&$\sigma_I$&$P_{\mu}$\\
(1)&(2)&(3)&(4)&(5)&(6)&(7)&(8)&(9)&(10)&(11)&(12)&(13)&(14)&(15)&(16)&(17)&(18) \\
&[h:m:s]&[d:m:s]&[pixel]&[pixel]&[mas/yr]&[mas/yr]&[mas/yr]&[mas/yr]&[mag]&[mag]&[mag]&[mag]&[mag]&[mag]&[mag]&[mag&[$\%$]\\
\hline
 1  & 16:47:23.05   &  -2:01:32.9 &  232.0060 &  222.3470 &   11.6618 &    0.4078 &    4.2128 &    0.4528 &   18.8176  &   0.0044 &   17.8204  &   0.0074 &   99.9999  &   0.9999 &   16.5937  &   0.0027  &  00.00 \\
 2  & 16:47:07.55   &  -2:01:32.1 & 1209.5560 &  228.9769 &   -0.6224 &    0.6910 &    0.4657 &    0.1974 &   17.5601  &   0.0011 &   16.7414  &   0.0003 &   99.9999  &   0.9999 &   15.6213  &   0.0033  &  97.45 \\
 3  & 16:46:41.64   &  -2:01:33.0 & 2843.5877 &  229.9219 &   -3.8028 &    0.3026 &   -3.4745 &    0.4292 &   17.9565  &   0.0039 &   17.2342  &   0.0030 &   99.9999  &   0.9999 &   16.2220  &   0.0044  &  00.00 \\
 4  & 16:47:03.98   &  -2:01:32.0 & 1434.2728 &  229.9849 &   -0.3477 &    0.3026 &    0.4399 &    0.7511 &   14.2924  &   0.0055 &   13.0543  &   0.0041 &   99.9999  &   0.9999 &   11.5623  &   0.0105  &  98.74 \\
 5  & 16:47:12.26   &  -2:01:31.1 &  912.3842 &  232.2896 &   -0.2361 &    0.4829 &    0.9142 &    0.6395 &   15.6701  &   0.0000 &   14.6978  &   0.0047 &   99.9999  &   0.9999 &   13.4579  &   0.0005  &  97.75 \\
 6  & 16:47:03.83   &  -2:01:26.1 & 1443.7522 &  254.8846 &    0.3176 &    0.0751 &   -0.0494 &    0.5387 &   16.1591  &   0.0049 &   15.2463  &   0.0047 &   99.9999  &   0.9999 &   14.0490  &   0.0077  &  99.65 \\
 7  & 16:47:14.38   &  -2:01:23.2 &  778.2754 &  264.8420 &    2.9659 &    0.2790 &    1.7212 &    0.2768 &   17.6799  &   0.0077 &   16.8928  &   0.0019 &   99.9999  &   0.9999 &   15.9671  &   0.0137  &  00.00 \\
 8  & 16:47:14.57   &  -2:01:21.9 &  766.4338 &  270.2744 &    0.3734 &    0.1416 &    0.1652 &    0.3520 &   16.7270  &   0.0022 &   15.8435  &   0.0047 &   99.9999  &   0.9999 &   14.7108  &   0.0005  &  99.65 \\
 9  & 16:47:10.68   &  -2:01:21.1 & 1011.7539 &  274.2942 &   -1.8306 &    0.8026 &   -2.8371 &    1.7212 &   18.5638  &   0.0033 &   17.8655  &   0.0261 &   99.9999  &   0.9999 &   16.9932  &   0.0028  &  43.65 \\
10  & 16:47:14.13   &  -2:01:20.6 &  794.4010 &  276.0190 &    0.3584 &    1.1782 &   -0.1438 &    0.5065 &   17.9872  &   0.0005 &   17.1587  &   0.0074 &   99.9999  &   0.9999 &   16.0792  &   0.0016  &  99.04 \\
11  & 16:47:10.98   &  -2:01:20.0 &  992.6786 &  279.0387 &    0.3305 &    0.6245 &   -0.3605 &    0.2597 &   18.4346  &   0.0050 &   17.6889  &   0.0003 &   99.9999  &   0.9999 &   16.6920  &   0.0049  &  99.39 \\
\hline
\label{cata}
\end{tabular}
}
\end{table*}
%\end{sidewaystable*}
%%%%%%%%%%%%%%%%%%%%%%%%%%%%%%%%%%%%%%%%%%%%%%%%%%%%%%%%%%%%%%%%%%%%%%%%%%%%%%%%%%%%%%%%%%%%%%%%%%%%%%%%%%%%%%%%%%%%%%%%%%%%%%%%%%%

\section{Conclusions}
\label{con}

In the present analysis, we have performed an astrometric and photometric investigation
on the globular cluster M12. Following conclusions can be drawn from the analysis.

\begin{enumerate}

\item A catalog of PMs and membership probabilities for 3725 stars in the
region of M12 has been presented.

\item PM catalog produced in the wide field region has been used to isolate cluster stars from field stars
and presented a clean CMD of M12.

\item The membership catalog has been used to ascertain the membership status of
variable stars, blue stragglers and $X$-ray sources reported earlier in the cluster's direction.

\end{enumerate}

\begin{acknowledgements}
We are grateful to the reviewer of this paper for very useful and constructive comments.
Devesh P. Sariya and Ing-Guey Jiang are supported by the grant from
the Ministry of Science and Technology (MOST), Taiwan.
The grant numbers are MOST 103-2112-M-007-020-MY3, MOST 104-2811-M-007-024, MOST 105-2811-M-007 -038,
MOST 105-2119-M-007 -029 -MY3, MOST 106-2112-M-007 -006 -MY3 and MOST 106-2811-M-007 -051.
This research is based on the observations with the MPG/ESO  2.2 m and ESO/VLT telescopes, located
at  La Silla  and  Paranal Observatory,  Chile,  under DDT  programs
163.O-0741(C), 085.A-9008(A) and the archive material.
This research used the facilities of the Canadian Astronomy Data Centre operated by the
National Research Council of Canada with the support of the Canadian Space Agency.
\end{acknowledgements}

\end{document}